\documentclass[aps,prb,twocolumn,showpacs,floatfix]{revtex4}

\usepackage{multirow}
\usepackage{graphicx}
\usepackage{amsmath}
\usepackage{subfigure}

\begin{document}

\title{Bound states of edge dislocations: The quantum dipole problem in two dimensions}

\author{K. Dasbiswas, D. Goswami, C.-D. Yoo and Alan T. Dorsey}

\affiliation{Department of Physics, University of Florida, P.O. Box
118440, Gainesville, FL 32611-8440}

\date{\today}

\begin{abstract}
We investigate bound state solutions of the 2D Schr\"odinger
equation with a dipole potential originating from the elastic
effects of a single edge dislocation. The knowledge of these states
could be useful for understanding a wide variety of physical
systems, including superfluid behavior along dislocations in solid
$^4$He. We present a review of the results obtained by previous
workers together with an improved variational estimate of the ground
state energy.  We then numerically solve the eigenvalue problem and
calculate the energy spectrum. In our dimensionless units, we find a
ground state energy of -0.139, which is lower than any previous
estimate. We also make successful contact with the behavior of the
energy spectrum as derived from semiclassical considerations.
\end{abstract}
 \maketitle

\section{Introduction}

There has been broad interest over the years in the physics of
solids containing dislocations. In addition to affecting the
mechanical properties of solids, the strain field associated with
dislocations binds charge carriers in metals, or solute impurities
in a generic solid.\cite{cottrell} As such, the presence of
dislocations has a significant effect on the transport, elastic and
superconducting properties of the solid. In this context, it is
important to know the spectrum of localized states due to a
dislocation.

In this article, we discuss the spectrum of bound states for an edge
dislocation. Within linear elasticity theory the  deformation
potential due to an edge dislocation is proportional to the stress
tensor or the divergence of the elastic displacement field.
Considering a straight edge dislocation, oriented along the
$z$-axis, within a continuum model, this potential is given by
\begin{equation}
V(r, \theta)= p\frac{\cos\theta}{r},
\end{equation}
where $p$ is the strength of the ``dipole'' potential, $r$ is the
distance from the dislocation axis and $\theta$ is the azimuthal
angle, both defined in the $x$-$y$ plane.\cite{cottrell} The dipole
moment $p$ depends on quantities such as the Fermi energy and the
lattice and elastic constants of the solid. In an electrostatics
context this potential can be realized as a dipole built by bringing
two infinite line charges of opposite sign close together. Here we
address the quantum dipole problem by considering the solution of
the corresponding two dimensional Schr\"odinger equation,
\begin{equation}
-\frac{\hbar^2}{2m} \nabla^2 \psi + p\frac{\cos\theta}{r} \psi = E
\psi. \label{schrodinger.eqn}
\end{equation}
For $p>0$ this potential is attractive for $x<0$ (thus, allowing for
bound states) and repulsive for $x>0$. It has parity in $y$; i.e.,
symmetry on reflection about the $x$-axis, which should be reflected
in the eigenfunctions as well. The solution of the Schr\"odinger
equation is complicated due to the non-central nature of the
potential.\cite{seeger} The potential being non-separable further
impairs the applicability of the WKB approximation.

We are particularly motivated by the supersolid
problem,\cite{balibar} and a possible interpretation of it which
considers superfluidity to exist not in the bulk of solid $^{4}$He
but along a network of dislocations.\cite{toner} We would like to
solve the full nonlinear Ginzburg-Landau (GL) theory for such a
system, for which we would first need to know the solution of the
linearized equation. The lowest eigenvalue of the linear equation is
actually a measure of the local enhancement in $T_c$ produced by a
dislocation. Further, the solution of the linear GL theory can be
used to affect a separation of the transverse degrees of freedom in
the full nonlinear time-dependent GL equation, resulting in a one
dimensional ``amplitude equation'' of superfluid density along the
dislocation. The effective dimensionless coupling constant of this
one dimensional theory is $g = \int dx dy {|\phi_0(x,y)|}^4$, where
${\phi_0(x,y)}$ the normalized ground state eigenfunction of the
linear GL equation. Our numerical solution of the linear equation
allows us to calculate this parameter which acts as an input to the
weakly nonlinear analysis.\cite{nextwork}

The problem of finding the ground state energy of the quantum dipole
problem has a long history, starting from the work of Landauer in
1954,\cite{landauer} who used a variational approach. Subsequent
authors used a variety of techniques for this estimate:
semiclassical \cite{lifschitz} or purely variational\cite{emtage,
slyusarev} methods, a combination of variational and perturbative
methods \cite{dubrovskii} or an expansion in terms of known basis
functions,\cite{nabutovskii, farvacque} but to our knowledge our
work the first to solve this using a direct numerical method. Some
prior works\cite{dubrovskii, slyusarev} have also studied the
spectrum of the bound eigenstates. The ground state energies
calculated in these works are shown in Table I, together with the
numerical value obtained in this paper.

\begin{table}[ht]
\caption{Summary of ground state energy estimates of the edge
dislocation potential. Energy is given in units of $2mp^2 /
\hbar^2$.}

\begin{tabular}{|c|c|}
\hline
       \multirow{2}{*}{{\bf References}} & {\bf Ground state}\\
        & {\bf estimate}\\
\hline \hline
        Landauer (1954) \cite{landauer} & -0.102\\
\hline
        Emtage (1967) \cite{emtage} & -0.117\\
\hline
        Nabutovskii and Shapiro (1977) \cite{nabutovskii}& -0.1014\\
\hline
        Slyusarev and Chishko (1984) \cite{slyusarev} & -0.1111\\
\hline
        Dubrovskii (1997) \cite{dubrovskii} & -0.1196\\
\hline
        Farvacque and Francois (2001) \cite{farvacque} & -0.1113\\
\hline
        Dorsey and Toner \cite{dorston} & -0.1199\\
\hline
                This work & -0.139\\
\hline

\end{tabular}

\end{table}

In the next section, we provide the details of our variational
calculation for the ground state energy, including our choice of a
suitable trial wave function. Next, we discuss the results and
technical details of the several numerical methods we have used to
calculate the eigenvalue spectrum, and their relative merits and
disadvantages. The methods involve diagonalization of the
Hamiltonian, carried out both in real space and in the basis of two
dimensional hydrogenic wave functions (in contrast with previous
calculations with different choices of basis expansions, e.g.
Refs.~\onlinecite{nabutovskii} and \onlinecite{farvacque}). Here we
also compare the results obtained using different methods, which are
found to agree in the essential features. In the final section, we
provide a semiclassical argument to justify our results together
with some discussion of the interesting properties of the classical
problem. The semiclassical result is found to extend to the lower
energy eigenstates as well.

\section{Variational Calculation}

Our initial approach to determine the ground state energy has been
variational because this can be carried out analytically and
provides a rough estimate which can then guide our more explicit
numerical solution. Given a normalized wave function $\psi(r,
\theta)$, we minimized the energy functional,
\begin{equation}
F[\psi,\psi^*] = \int d^2 x \left( \frac{\hbar^2}{2m} |\nabla
\psi|^2 + p\frac{\cos\theta}{r} |\psi|^2 \right). \label{functional}
\end{equation}
This functional has its extrema at the solutions of the
Schr\"odinger equation, Eq.~(\ref{schrodinger.eqn}). Note that the
length and energy scales which emerge from Eq.~(\ref{functional})
(or the Schr\"odinger equation) for this problem are $\hbar^2/2mp$
and $2mp^2/\hbar^2$. In dimensionless variables, the normalized
trial wave function used in our calculation is
\begin{eqnarray}
\psi(r,\theta)&=&\frac{2AB}{C\sqrt\pi}\frac{(1-r/BC)}{\sqrt{(3-4B+2B^2)}}\exp\bigg(-\frac{r}{C}\bigg)\nonumber
\\
&& -\frac{\sqrt{1-A^2}}{C^2}\sqrt{\frac{8}{3\pi}}r\cos\theta
\exp\bigg(-\frac{r}{C}\bigg),
\end{eqnarray}
where $A$, $B$ and $C$ are variational parameters. We choose the
trial wave function so as to account for the anisotropy of the
potential. Further, the asymptotic behavior of the potential is
captured by the exponentially decaying factors. The minimum
expectation value of the energy occurs when $A=0.803, B= -0.774$ and
$C=2.14$ with a ground state energy of $-0.1199$ which was found by
Dorsey and Toner.\cite{dorston} This value is  2.5\% lower than the
previous lowest variational estimate ($-0.1196$) obtained by
Dubrovskii.\cite{dubrovskii} In addition, by using this normalized
trial wave function as the $\phi_0(x,y)$ we find the parameter $g =
\int dx dy {|\phi_0(x,y)|}^4 = 0.017$.

\section{Numerical Methods}

A detailed numerical solution of the two-dimensional Schr\"odinger
equation with the dipole potential, Eq.~(\ref{schrodinger.eqn}), is
likely to provide more accurate ground state eigenvalues in addition
to determining the rest of the bound state eigenvalues and
corresponding wavefunctions. We do this both by a real space
diagonalization, where the Schr\"odinger equation is discretized on
a square grid, and by expanding in the basis of the eigenfunctions
of the two-dimensional Coulomb potential problem. Two special
features of this dipole potential make it a numerically difficult
problem: the singularity at the origin, and the long range behavior
of the potential. It is expected that the Coulomb wavefunctions
would be better suited to capturing this long range behavior, and
convergence would consequently be faster. Our results show that the
Coulomb basis method is more accurate for the higher bound states
(which are expected to extend more in space), as the real space
methods are limited by size issues. However, the real space method
works better for the ground state.

\subsection{Real Space Diagonalization Method}

For numerical purposes the Schr\"odinger equation is converted to a
difference equation on a square grid of spacing $h$, with the
Laplacian approximated by its five-point finite difference
form,\cite{numeric} resulting in a block tridiagonal matrix of size
$N^2\times N^2$, where the grid has dimensions of $N\times N$. Each
diagonal element corresponds to a grid point and has values of
$4/h^{2} + V(x,y)$, whereas the nonzero offdiagonal elements all
equal $-1/h^{2}$. The matrix is thus very large but sparse. We use
three different numerical methods to diagonalize this matrix: the
biconjugate gradient method,\cite{bcg} the Jacobi-Davidson algorithm
\cite{j-d} and Arnoldi-Lanczos algorithm,\cite{lanczos} with the
latter two being more suited to large sparse matrices whose extreme
eigenvalues are required. We use freely available open source
packages (JADAMILU \cite{jadamilu} and ARPACK \cite{arpack}) written
in FORTRAN for both.  All three approaches are  projective Krylov
subspace methods, which rely on repeated matrix-vector
multiplications while searching for approximations to the required
eigenvector in a subspace of increasing dimensions.
Reference~\onlinecite{jadapaper} provides a concise introduction to
the Jacobi-Davidson method, together with comparisons to other
similar methods. The implicitly restarted Arnoldi package (ARPACK)
is described in great detail in Ref.~\onlinecite{arpmanual}. Some
general issues about the real space diagonalization as well as some
specific features of the three methods used for it are discussed
below.

The accuracy of the real space diagonalization methods is controlled
by two main parameters: the grid spacing $h$ and the total size of
the grid, which is given by $Nh$.  The finite difference
approximation together with the rapid variation of the potential
near the origin imply that the solution of the partial differential
equation would be more accurate for a smaller grid spacing. We work
with open boundary conditions, which means that a bound state
wavefunction could be correctly captured only if the total size of
the grid were to be greater than the natural decay length of the
wavefunction. In other words, the eigenstate has to be given enough
space to relax. This limits the number of bound states we can
calculate accurately because a large grid size together with small
grid spacings calls for a large number of grid points, thus
quadratically increasing the size of the matrix to be diagonalized.
Computational resources as well as the limitations of the algorithms
themselves place an effective upper bound on the size of a
diagonalizable matrix. We experimented to find that a
$10^6\times10^6$ size sparse matrix was about the maximum that could
be diagonalized with our computational resources.


The origin of the square grid is symmetrically offset in both $x$
and $y$ directions to avoid the $1/r$ singularity. We tested first
the accuracy of the real space techniques for the case of the two
dimensional Coulomb potential, the spectrum of which is completely
known.\cite{yang91} We observe that for various lattice sizes the
biconjugate method captures at most the first four states whereas
the Jacobi-Davidson method returns 20 eigenstates. The eigenvalues
obtained from both methods are accurate to within 2\% of the exact
values.\cite{yang91}

We have applied the biconjugate method to the edge dislocation
potential for various lattice sizes, varying from 10$\times$10 to
600$\times$600. The number of eigenstates captured increases with
the size of the lattice, as expected. The ground state energy is
observed to vary from -0.134 to -0.142. We also observe that for the
number of grid points exceeding $N=2000$ we encounter a numerical
instability due to the accumulation of roundoff errors. For the
largest real space grid size of 600$\times$600 ($N = 1200, h = 0.5$)
we obtain seven eigenstates with a ground state energy of -0.1395.

The ground state energy from the Jacobi-Davidson method, employed
for the same lattice size gives -0.1395, which matches well with our
expectations from the variational calculation. We are able to obtain
$20$ bound state eigenvalues in this method using $N=1000, h= 0.5$.
It is checked that the low-lying eigenvalues are not very sensitive
to values of $h$ in this regime, so a relatively large value of
$0.5$ serves our purpose.

The Arnoldi-Lanczos method yields very similar eigenvalues. It takes
more time and memory resources to converge but can calculate more
eigenvalues, with greater accuracy for the higher excited states. It
provides $30$ bound state eigenvalues for the same set of lattice
parameters as the above. Finally, after calculating the ground state
wave function we find that the coupling constant $g=0.0194$,
slightly larger than the variational estimate of $g=0.017$.

\begin{figure}[t]
    \centering
        \includegraphics[width= 3in]{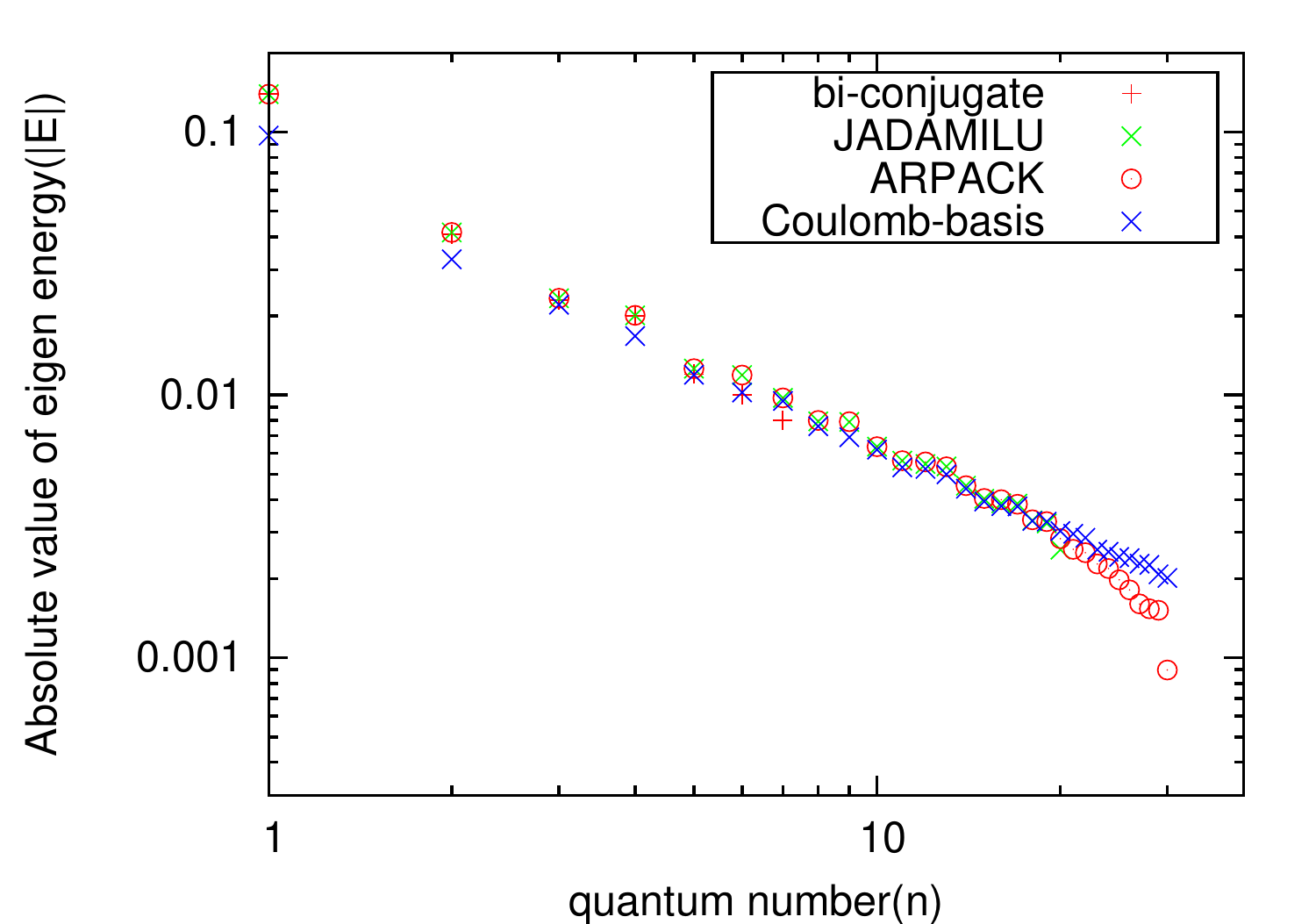}
        \caption{(Color online) Comparison of eigenvalues obtained from different methods. (The plot is on a log-log scale.) \label{fig1}}
\end{figure}

\begin{table}[t]
\caption{Comparison of first few energy eigenvalues obtained from
different methods. Energy units: $2mp^2 / \hbar^2$. $n$ indicates
quantum number of the state. \label{table2}}
\begin{tabular}{|c|c|c|c|c|}
\hline
  \multirow{2}{*}{$n$}     & \multirow{2}{*}{{\bf biconjugate}}   & {\bf Jacobi-} & {\bf Arnoldi-}\ & {\bf Coulomb}\\
      &    & {\bf Davidson} & {\bf Lanczos}\ & {\bf basis}\\
\hline \hline
 1   & -0.14 & -0.13954  & -0.13952   & -0.09697\\
\hline
 2 & -0.041 & -0.041480 & -0.041478  &  -0.03281\\
\hline
 3 & -0.023 & -0.023314 &  -0.023314 & -0.022067 \\
\hline
 4 & -0.02 & -0.020086 &  -0.020086 &  -0.016744 \\
\hline
 5 & -0.012 & -0.012592 &  -0.012594 &  -0.011944 \\
\hline
\end{tabular}

\end{table}

\subsection{Coulomb Basis Method}

We also calculate the spectrum numerically by using the linear
variational method with the basis of  the 2D hydrogen atom wave
functions \cite{yang91}. There are two advantages of this method
over the real space diagonalization methods. First, the linear
variational method is capable of capturing more excited states
because the number of calculated bound states is not limited by the
size of the real space grid but by the number of long-range basis
functions. Second, the singularity at the origin of the edge
dislocation potential does not pose a problem anymore because
elements of the Hamiltonian matrix become integrable.

Now we calculate the elements of the Hamiltonian matrix with a 2D
edge dislocation potential. The Schr\"odinger equation with the 2D
Coulomb potential is analytically worked out in
Ref.~\onlinecite{yang91}. The normalized wave functions of a 2D
hydrogen atom are given by
\begin{equation}
\psi^\text{H}_{n,l}(r,\theta)= \sqrt{\frac{1}{\pi}} R_{n,l}(r)
\times
\begin{cases}
\cos(l \theta) &\text{for $1 \leq l \leq n$},
\\
\frac{1}{\sqrt{2}} &\text{for $l = 0$},
\\
\sin(l \theta) &\text{for $-n \leq l \leq -1$},
\end{cases}
\end{equation}
where
\begin{equation}
\begin{split}
R_{n,l}& (r) =
\frac{\beta_n}{(2|l|)!}\sqrt{\frac{(n+|l|-1)!}{(2n-1)(n-|l|-1)!}}
(\beta_n r)^{|l|}
\\
&\times\exp\left(-\frac{\beta_n r}{2}\right)
{_1}F_1(-n+|l|+1,2|l|+1,\beta_n r),
\end{split}
\end{equation}
with $\beta_{n} = 2/(2n-1)$ and ${_1}F_1$ being the confluent
hypergeometric function. The elements of the Hamiltonian with the 2D
dipole potential are
\begin{equation}
\begin{split}
\left< \psi^\text{H}_{n_1, l_1} | -\nabla^2 | \psi^\text{H}_{n_2,
l_2} \right> &= \delta_{l_1, l_2} \int_0^\infty d r \; \left( 1 -
\frac{\beta_{n_2}^2}{4}r\right)
\\
& \times R_{n_1,l_1} (r)  R_{n_2,l_2} (r),
\end{split}
\end{equation}
\begin{equation}
\left< \psi^\text{H}_{n_1, l_1} \bigg| \frac{\cos\theta}{r} \bigg|
\psi^\text{H}_{n_2, l_2} \right> = \tilde{V} \int_0^\infty dr \;
R_{n_1,l_1} (r)  R_{n_2,l_2} (r),
\end{equation}
where $\tilde{V}=\delta_{l_1, l_2\pm1} / 2$ if both $l_1$ and $l_2$
are less or greater than 0, or $\tilde{V}=1 / \sqrt{2}$ if $l_1$ is
0 and $l_2$ positive or vice versa. The spectra are obtained for
several total numbers of basis functions $N_\text{basis}$. Due to
the numerical precision in calculating elements of the Hamiltonian
matrix $N_\text{basis}$ cannot be increased to more than 400. For
$N_\text{basis}=400$ we obtain about 149 bound states and the ground
state energy of -0.0969. In order to improve the ground state
energy, we introduce an additional decaying parameter in the basis
functions, and optimize the energy levels for a certain value of
this parameter. With the decaying parameter we obtain the best
variational estimate for the ground state energy of -0.1257 for
$N_\text{basis}=400$.

We show the first twenty eigenvalues obtained from different methods
in Fig.~\ref{fig1} and the first five representative eigenvalues in
Table~\ref{table2}. As mentioned earlier, the real space
diagonalization methods provides a best estimate of the ground state
energy whereas the Coulomb basis method is more suitable for higher
excited states. The eigenvalues of both the Coulomb basis method and
the real space diagonalization methods are found to match each other
for excited states, and then they begin to deviate again (see
Fig.~\ref{fig2}). This can be understood by the fact that the extent
of wave functions of the 2D edge dislocation potential does not
always increase as one goes to higher excited states -- the wave
functions of some excited states extend less than those of lower
energy. Therefore, there are intermediate bound states that are to
be missed in the real space calculation because the size of grid
used in calculation is not large enough to capture them. For
example, we find four more bound states with the Coulomb basis
calculation between the 18$^\text{th}$ and 19$^\text{th}$ excited
states as calculated from the real space diagonalization method.
This feature also explains the abrupt increase of the eigenvalue of
the 19$^\text{th}$ state calculated by using the Arnoldi-Lanczos
method (ARPACK routine) in Fig.~\ref{fig2}.

\section{Semiclassical Analysis}

\begin{figure}[t]
  \centering
  \includegraphics[width = 3in]{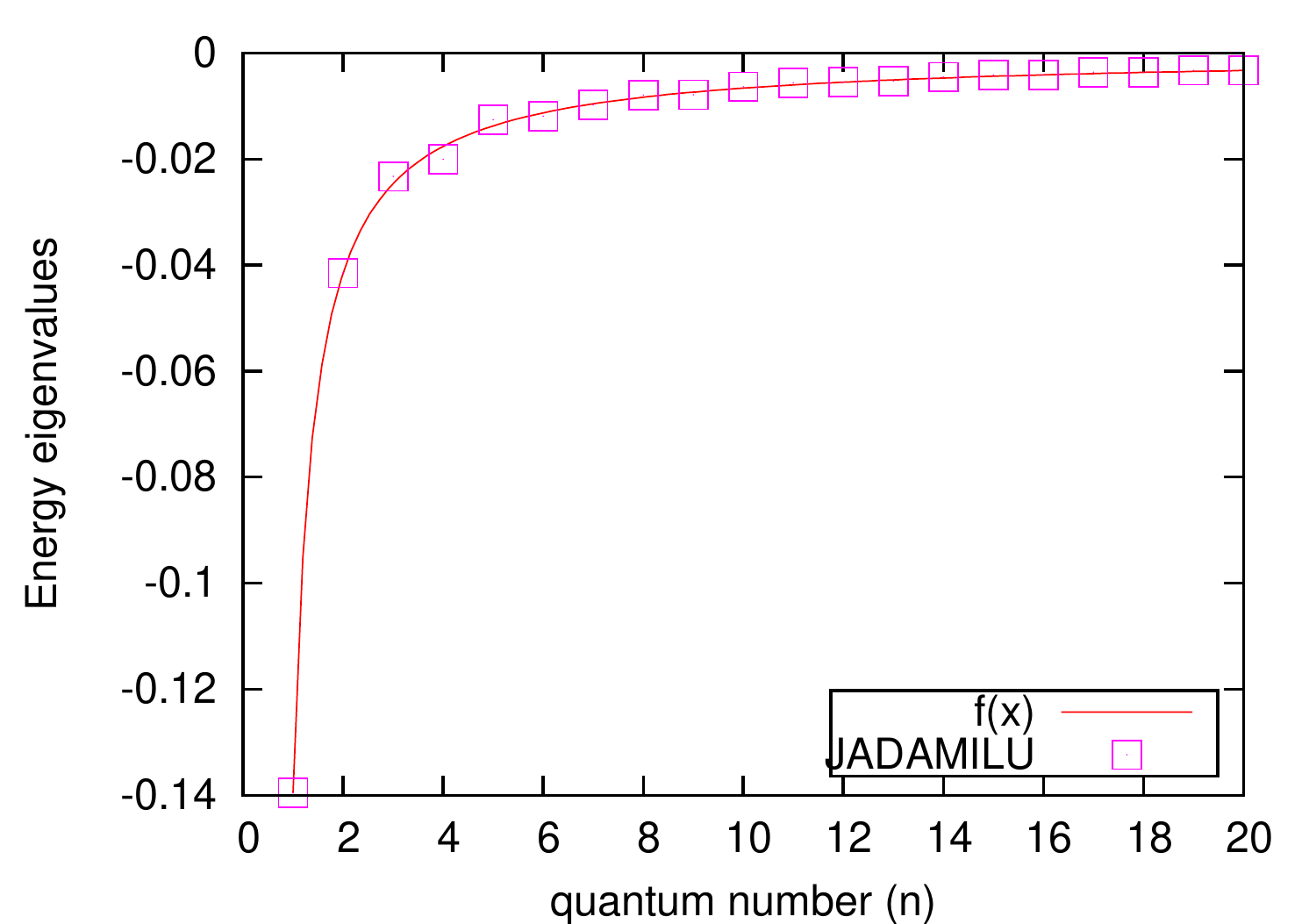}
  \caption{(Color online) Fit for the eigenvalue spectrum obtained from JADAMILU using $f(x)=a(x-b)^c$. Fit values are $-0.06, 0.61$ and $0.96$ for $a, b$ and $c$ respectively. \label{fig2}}
\end{figure}

It is usually insightful to consider the semiclassical solution of a
quantum mechanics problem, since the higher energy eigenstates tend
to approach classical behavior. A semiclassical estimate of the
energy spectrum has been provided in Ref.~\onlinecite{lifschitz}.
Here the total number of eigenstates up to a value of energy $E$ is
proportional to the volume occupied by the system in the classical
phase space. This is expressed by Weyl's theorem \cite{Weyl} :
\begin{equation}
n(E) = \frac{A}{4\pi} \frac{2m|E|}{\hbar^2} +
\mathcal{O}\left(\sqrt{\frac{\hbar^2}{2mp^2}|E|}\right),
\end{equation}
where $A$ is the classically accessible area in real space and $|E|$
the absolute value of energy of the state. The higher order
corrections can be shown to be less important for higher excited
states, which is where the semiclassical picture applies. To find
$A$, we need the classical turning points for this potential
determined by setting $E=V(r,\theta)$. Then the accessible area is
the interior of a circle given by $(x- \frac{p}{2E})^2 + y^2 =
(\frac{p}{2E})^2 $, with area $A=\pi(p/2E)^2$. Therefore, we obtain
(writing the nondimensionalized energy in our system of units as
$\epsilon$):
\begin{equation}
n(\epsilon) = -\frac{1}{16\epsilon},
\end{equation}
where $n$ is the quantum number of the eigenstate, and $\epsilon$
the corresponding energy. Note that the density of states
$dn/d\epsilon$ scales as $1 /\epsilon^2$.

To check this result we fit the numerical spectrum with the
following functional form:
\begin{equation}
\epsilon(n) =  a(n-b)^c,
\end{equation}
with the fitting parameters having values $a = -0.06, b=0.5,
c=-0.98$, each correct to within $5\%$. (Since we are dealing with
bound states here, all the energy eigenvalues are negative, and the
higher excited states have lower absolute eigenvalues.) We show the
fit to the spectrum obtained from JADAMILU routine in
Fig.~\ref{fig2}. The semiclassically derived dependence is found to
closely match with the fit for numerically calculated energy
eigenstates, except for the $b=0.5$ factor. In the limit of large
$n$ values i.e, higher excited states, the fit relation tends to the
semiclassical result as expected.

\begin{figure}[t]
  \begin{center}
    \subfigure[Eigenfunction for the ground state]{\label{fig:a}\includegraphics[width = 2 in]{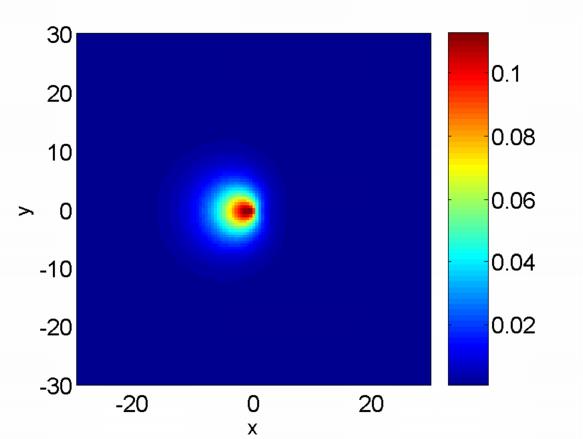}}
    \subfigure[Eigenfunction for the 1st excited state]{\label{fig:b}\includegraphics[width = 2 in]{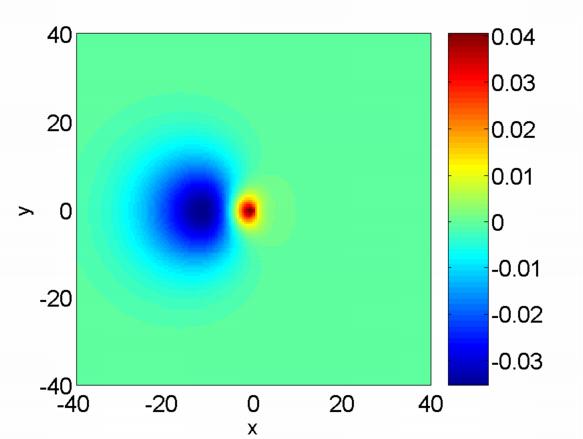}} \\
  \end{center}
  \caption{(Color online) Eigenfunctions of the first two bound states}
 \label{fig}
\end{figure}
The classical trajectories for this potential bear the signature of
chaotic dynamics showing space-filling nature and strong dependence
on initial conditions. However, for reasons not yet clear to us,
they are not ergodic, filling up only a wedge-shaped region in real
space instead of the full classically allowed circle. The quantum
mechanical probability density as calculated from the eigenfunctions
also exhibits such wedge-shaped regions.  Some sample wavefunctions
obtained from our numerical calculations have been included in
Fig.~\ref{fig}. The parity of the potential shows up in the
wavefunctions being either symmetric or antisymmetric about the
$x$-axis, although states of such ``odd'' and ``even'' parity do not
always alternate. Also, as mentioned earlier, the spatial extent of
the wavefunctions does not scale monotonically with quantum number.
We do not have any satisfactory explanation yet for these irregular
features.

\section{Summary}

In conclusion, we have investigated the longstanding quantum problem
of a two-dimensional dipole potential. The wave functions and the
spectrum are calculated by solving  the Schr\"odinger equation with
the 2D dipole potential numerically, and also, in the case of the
ground state, variationally. We find that the results obtained from
the different methods are consistent and compare favorably with
previous estimates in the literature. We also discover a simple
pattern in the spectrum, ($n \propto \epsilon^{-1}$), which can be
justified from semiclassical considerations. Certain features of the
spectrum and wave functions are yet to be explained and might
provide scope for future investigation. For example, the statistics
of the level spacings could possibly be a signature of quantum
chaos. We hope to extend our work to studying dislocation-induced
superfluidity as a model of $^4$He supersolid.\cite{nextwork} In
such a model, the linearized GL equation is isomorphic to the
Schr\"odinger equation, Eq.~(\ref{schrodinger.eqn}), and the ground
state energy and its wave function determined in this work provides
an input to obtain the coupling constant $g$ and the superfluid
transition temperature modified by the presence of dislocations.


\begin{acknowledgments}
We are grateful to H. B. Thacker for kindly providing us with a 2D
Schr\"odinger equation solver based on the biconjugate gradient
method, J. Toner for useful discussions and ideas, C. Taylor for
providing computational support, and S. Bera and A. Kemper for
important suggestions regarding numerical methods. The authors
acknowledge the University of Florida High-Performance Computing
Center for providing computational resources and support that have
contributed to the research results reported within this paper. This
work was supported by NSF Grant No. DMR-0705690.
\end{acknowledgments}


\end{document}